\documentclass[%
aip,
rsi,
reprint,
amsmath, amssymb
]{revtex4-1}

\usepackage{graphicx}
\usepackage{dcolumn}
\usepackage{amsmath}
\usepackage[mathlines]{lineno}

\usepackage[utf8]{inputenc}
\usepackage[T1]{fontenc}
\usepackage{mathptmx}
\usepackage{etoolbox}
\usepackage{xcolor}
\makeatletter
\def\@email#1#2{%
 \endgroup
 \patchcmd{\titleblock@produce}
  {\frontmatter@RRAPformat}
  {\frontmatter@RRAPformat{\produce@RRAP{*#1\href{mailto:#2}{#2}}}\frontmatter@RRAPformat}
  {}{}
}%
\makeatother
\begin{document}

\title{Identifying environmentally induced calibration changes in cryogenic RF axion detector systems using Deep Neural Networks}
\author{Andrew Engel}
\affiliation{ 
Pacific Northwest National Laboratory, Richland, WA, USA}
\affiliation{Department of Physics, Ohio State University, Columbus, OH, USA}
\affiliation{Center for Cosmology and AstroParticle Physics (CCAPP)\\, Ohio State University, Columbus, OH, USA}

\author{Thomas Braine}%
 \email{thomas.braine@PNNL.GOV}
\author{Christian Boutan}
\affiliation{ 
Pacific Northwest National Laboratory, Richland, WA, USA
}%

\date{\today}

\begin{abstract}
The axion is a compelling hypothetical particle that could account for the dark matter in our universe, while simultaneously explaining why quark interactions within the neutron do not appear to give rise to an electric dipole moment. The most sensitive axion detection technique in the 1–10 GHz frequency range makes use of the axion-photon coupling and is called the “axion haloscope”. Within a high Q cavity immersed in a strong magnetic field, axions are converted to microwave photons. As searches scan up in axion mass, towards the parameter space favored by theoretical predictions, individual cavity sizes decrease in order to achieve higher frequencies. This shrinking cavity volume translates directly to a loss in signal-to-noise, motivating the plan to replace individual cavity detectors with arrays of cavities. When the transition from one to (N) multiple cavities occurs, haloscope searches are anticipated to become much more complicated to operate: requiring N times as many measurements but also the new requirement that N detectors function in lock step. To offset this anticipated increase in detector complexity, we aim to develop new tools for diagnosing low temperature RF experiments using neural networks for pattern recognition. Current haloscope experiments monitor the scattering parameters of their RF receiver for periodically measuring cavity quality factor and coupling. However off-resonant data remains relatively useless.

In this paper, we ask whether the off resonant information contained in these VNA scans could be used to diagnose equipment failures/anomalies and measure physical conditions (e.g., temperatures and ambient magnetic field strengths). We demonstrate a proof-of-concept that AI techniques can help manage the overall complexity of an axion haloscope search for operators. 
\end{abstract}

\maketitle

\section{\label{sec:intro}Introduction}
There is significant cosmological and astrophysical evidence for dark matter, matter that only interacts by gravitation appreciably, contributing about 85\% of the total mass in our universe \cite{Zwicky1933,VRubin1970,PlanckCollab2021}. Identification and classification of this dark matter remains one of physics greatest outstanding problems. While there are a number of hypothetical candidates for the dark matter particle, such weakly interacting massive particles (WIMPs) \cite{LUXZEP2023,XenonnT2023,PandaX2021} or sterile neutrinos \cite{SterileNeutrinos1,SterileNeutrinos2,NuStar}, the invisible QCD axion is particularly well-motivated because it not only solves the dark matter problem, but yet another problem with strong nuclear theory, the strong CP problem \cite{PecceiQuinn1977,Weinberg1978,Wilczek1978}. 
The most sensitive experimental technique to detect axions thus far has been the axion haloscope, first proposed in Ref. \cite{Sikivie1983}; The Axion Dark Matter eXperiment (ADMX) pioneered this detector technology and was the first to reach discovery potential sensitivity\cite{ADMX2001,ADMX2004,ADMX2010,ADMX2018,ADMX2020,ADMX2021}. Future, higher frequency detectors are plagued by scalability problems, essentially requiring large arrays of smaller haloscopes that can signal combine to increase sensitivity 
Such a complex system will require increased diagnostic information to monitor the performance and evaluation of its sensitivity at any given time. This section will introduce the axion, the axion haloscope, and the scalability problems of future detectors.
\subsection{\label{sec:axions} Axions}
The axion was first proposed as a consequence of $U(1)$ symmetry introduced to force CP conservation in the strong nuclear force. According to the current standard model, the strong nuclear force can be CP-violating, yet experimental measurements place an very small upper limit on the CP violating phase, $\theta<5 \times 10^{-11}$; This limit is most commonly set by measuring the upper bound on the neutron electric dipole moment, which is consistent with a zero value \cite{nEDM,axionTheta}. Without a mechanism for CP conservation, this constitutes a fine-tuning problem.

By introducing a global axial $U(1)_{PQ}$ symmetry that undergoes spontaneous symmetry breaking at a very high temperature, $T_{PQ}$, in the early universe, the CP-violating phase would naturally relax to a zero value, $ \theta =0$ \cite{PecceiQuinn1977}. Additionally, a pseudo Nambu-Goldstone boson, the QCD axion, would be produced as a result of this symmetry-breaking \cite{Weinberg1978, Wilczek1978}.

These axion particles would be massive, yet very weakly coupled to photons, making them 'invisible', therefore an ideal dark matter candidate. Through a process called vacuum re-alignment, enough invisible axions could be produced to make up the entirety of the dark matter \cite{axionReviewSikivie2021}. If the transition temperature, $T_{PQ}$, is lower than the reheating temperature after cosmological inflation, these axions would most likely have a mass between $\mathcal{O}(1\, \mu eV)$ and $\mathcal{O}(1\, meV)$ \cite{axionTheta}. Two benchmark models, describe the relation between the axion mass and its weak coupling to photons: the Kim-Shifman-Vainshtein-Zakharov (KSVZ) \cite{KSVZ1,KSVZ2} and Dine-Fischler-Srednicki-Zhitnitsky \cite{DFSZ1,DFSZ2} (DFSZ) models. In order to discover these axions, one wants a detector that can probe this entire likley range of axion masses, with a sensitivity to the KSVZ and DFSZ models; The ADMX haloscope was the first of its kind to do this. 

\subsection{Axion Haloscopes}
\label{sec:axion_haloscopes}
The axion haloscope was first proposed by Pierre Sikivie in 1983 \cite{Sikivie1983}, to look for dark matter axions via the inverse Primakoff effect \cite{Primakoff}; axions are stimulated to convert to photons via a strong magnetic field. The photons will have an energy equal to the axion's rest mass and kinetic energy, but because dark matter is cold and non-relativistic, the kinetic term will be negligible. Based on the likely axion mass range, $\mathcal{O}(1\, \mu eV)$ and $\mathcal{O}(1\, meV)$, this would correspond roughly to frequencies of 0.25-250 GHz.
\par A microwave cavity resonator with a mode tuned to the corresponding photon frequency will resonantly enhance the axion photon power by the cavity quality factor. This signal can then be read out by an antenna sampling the cavity power, with an ultra low-noise RF receiver. Because the axion mass is unknown, cavity tuning structures must be used to adjust the resonant frequency of the search mode, most commonly the $TM_{010}$ mode, because its structure maximizes the efficiency of axion photon power.
\par
\begin{figure*}[!ht]
    \centering
    \includegraphics[scale=0.7]{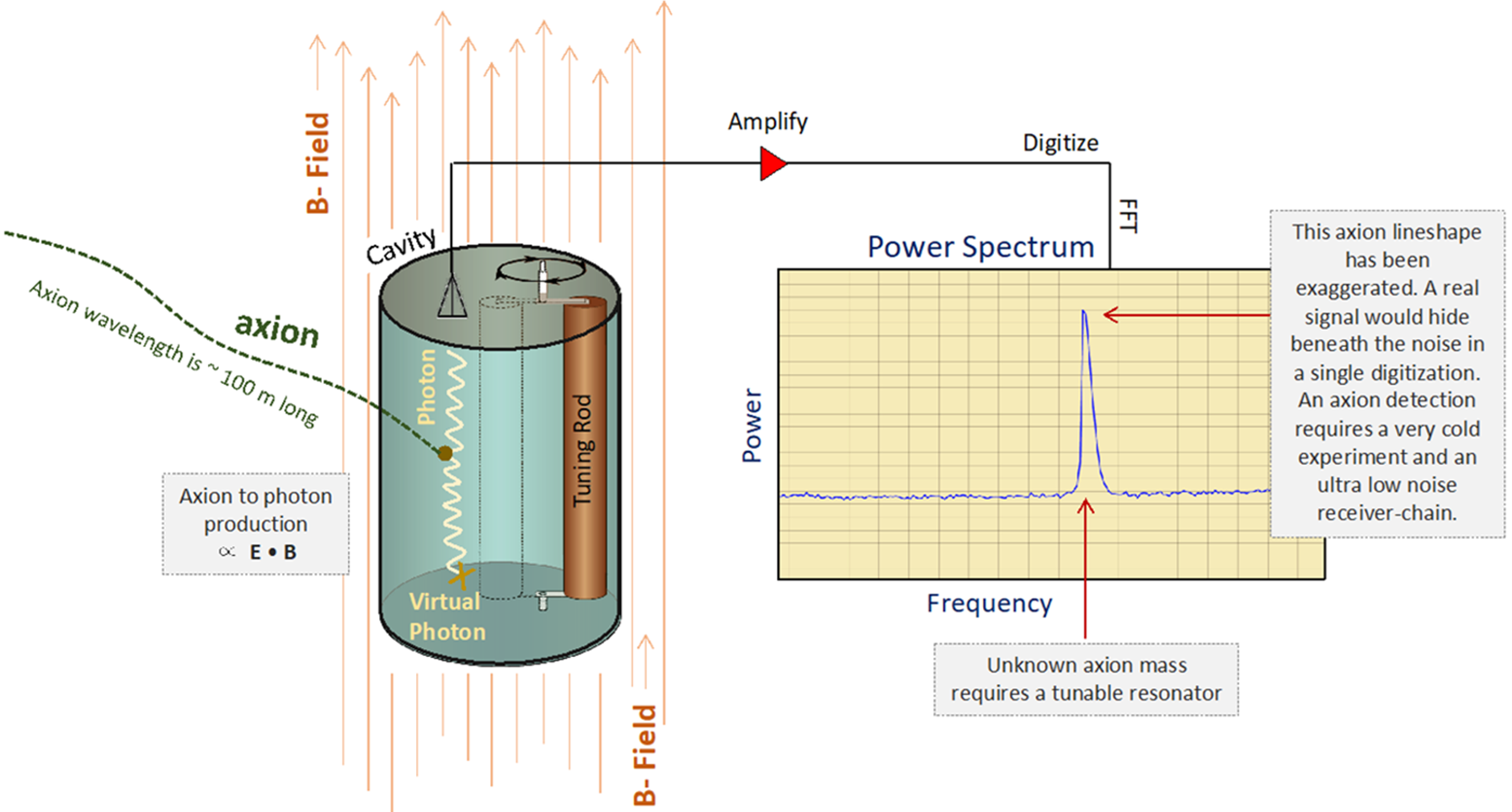} 
    \caption{A simplified diagram of a Axion Haloscope \cite{BoutanThesis}. Axions are converted to photons via magnetic stimulation, and are resonantly enhanced by a microwave cavity. Power from the cavity is sampled via an antenna, amplified by an ultra-low noise receiver, and digitized. An axion signal would manifest as a small, persistent power excess above the thermal background energy spectrum.}
    \label{fig:Haloscope}
\end{figure*}

 The dominant background in such an experiment is thermal noise, therefore the detector is cooled to milli-Kelvin temperatures via a helium dilution refrigerator. Quantum-noise-limited amplifiers are used within the receiver chain to further minimize added noise to the signal as it is read out. Other RF components such as circulators and attenuators are incorporated to minimize reflections and added noise along the signal path. The ultimate goal is to detect a persistent power excess signal within a Fourier transform about the resonant region; This is all pictured in the diagram shown in Figure \ref{fig:Haloscope}.
\par Diagnostic information on how the detector is performing is crucial to determining its sensitivity in any given energy spectrum digitization. Temperature sensors monitor the various temperature stages and components for excess heat. Periodic measurements of the cavity quality factor, antenna coupling, and system noise temperature allow the operator to drive the experiment at the desired power sensitivity as the cavity tunes through its frequency range. Because of this, up to 27\% of a given data-taking cycle maybe spent not taking axion search data, but characterizing and optimizing the receiver performance \cite{ADMX1BAnalysisDetails}. Even with that information, current detectors still have blind spots for operators; one example of this would be the exposure of sensitive electronics to stray magnetic fields, which are currently protected by a 'bucking' magnet that cancels out the main magnet field. One may ask is there a way to extract such information from the existing diagnostic data that is already taken through the use of machine learning and artificial intelligence.

\subsection{Future Detectors}
\label{sec:future_haloscopes}
Future axion haloscopes face many engineering challenges. Thus far, ADMX has excluded axions, moving up in frequency, from roughly 0.6-1.0 GHz at DFSZ sensitivity \cite{ADMX2018,ADMX2020,ADMX2021}, with plans to cover up to 2.0 GHz. With each data run, the cavity tuning rods must be enlarged or the cavity radius decreased in order to increase the resonant frequency of the cavity mode; This comes with the detriment of detection volume loss. This volume loss translates to a slow down of $\frac{df}{dt}\propto f^{-4}$ in scan speed if one keeps using a single cavity.
\par Multi-cavity arrays can circumvent this problem by signal combining smaller, high frequency cavities to maintain volume. Above 1.3 GHz, ADMX plans to transition to a 4-cavity array to cover frequencies out to 2 GHz \cite{4cav_boutan_aps}; there are also plans for an 18-cavity array to cover the 2-4 GHz region still in development. 
\par These arrays, however, will be increasingly complex, and require diagnostics and controls that keep the cavities operating together efficiently. With each cavity comes more antennas, circulators, tuning motors, and amplifiers that the operator needs monitored in case of failure. Resonant frequencies between cavities need to be matched, quality factors maintained to similar value, and low-noise receivers optimized to similar noise temperatures in order for signal combining to be effective and efficient. With dead time already significant in single cavity systems, there is not much time to spare acquiring more diagnostic data in multi-cavity systems.
\par Machine learning and Artificial intelligence could be a saving technology for the future haloscope operator. For instance, perhaps a neural network could be used to detect that single faulty RF circulator in array of potentially 50 circulators, using existing diagnostic data, that a human operator would never be able to detect. Being a scanning experiment that takes year-long data runs, AI could be used for forecasting potential slow-downs due to the combination of a myriad of factors that a human operator would not foresee; for instance, the system of receiver chains might experience an overall increase in system noise temperature that would not be predicted looking at the individual receiver performance. Large language models could be used to ease the human requirements of experimental monitoring, providing alerts when existing sensors go outside their expected range. 
\par This paper specifically aims to show how off-resonant scattering parameters, data already collected in current haloscope experiments, can be fed into a neural network to diagnose anomalies and identify physical changes in a single RF component, specifically a RF circulator.

\section{Background}
This section aims to give the background needed to understand the three rudimentary machine learning experiments that follow in the next section. Section \ref{sec:scattering_params} will explain scattering parameters in RF networks. Section \ref{sec:CryoRF} will point out the challenges of cryogenic RF environments inside an axion haloscope. Section \ref{sec:ANNs} stresses the artifical neural network (ANN) as solution for inhuman pattern recognition in otherwise noisy data, and finally section \ref{sec:extractdiagnostics} proposes the application of concern for the remainder of the paper, using ANNs to extract diagnostics from scattering parameter data.

\subsection{Scattering Parameters}
\label{sec:scattering_params}
Scattering parameters, also know as S-parameters, can be used to characterize the behavior of RF networks when stimulated by electrical signals. An RF network can be considered an arrangement of RF devices with some number of open ports that electrical signals can be inputted or measured from. For a given complex voltage signal at a pure tone frequency, $\tilde{V}_{in}(f)$, inputted into port $j$ of the network and resultant voltage signal, $\tilde{V}_{out}(f)$, measured from port $i$, the S-parameter can be defined:
\begin{equation}
    S_{ij}(f)=\frac{\tilde{V}_{out}(f)}{\tilde{V}_{in}(f)}
\end{equation}
Popularly, the S-parameters are thought of as a ratio between input and output powers, which is true in the case of the magnitude of a complex S-parameter, but not the case for the phase component, so the voltage definition is more accurate. For a N-port network this can be re-written as the S-matrix equation:
\begin{equation}
    \tilde{V}_{out}=\mathbf{S}\tilde{V}_{in}
\end{equation}
In the case of a 1 port network, only the reflected power off the input port can be measured, $S_{11}$. In the case of a 2 port network, both transmitted ($i\neq j)$ and reflected power ($i = j)$ through each port can be measured, forming a 2x2 matrix. Higher N-port networks can similarly be described by more reflection and transmission measurements. 
\par S-parameters are most typically measured by a vector network analyzer (VNA). The network analyzer is essentially two devices working in conjunction with one another: a signal generator makes known complex voltage signals to input into a given network port, and a spectrum analyzer measures the resultant signal at the output port. By inputting a fixed power, swept frequency signal by the signal generator and normalizing the complex power spectrum measured at the spectrum analyzer by the input power, $S_{ij}(f)$ is measured as a function of frequency. The 'vector' in VNA refers to its ability to measure both the magnitude and phase of these signals. Typically, the S-parameter values are referred to in relative units of decibels (dB), because it is defined as a ratio between input and output signals; linear units are just as valid however. VNAs typically operate from 0-20 GHz, at a variety of input powers (-80 to +20 dBm).
\par A variety of devices can be characterized by their expected S-parameters \cite{Caspers2012}. In the case of 1-port, there only exists 3 possible devices: a RF short where $-1>|S_{11}|>0$, a terminator where $0<|S_{11}|<1$, and an active reflection amplifier, $S_{11}>1$, with values being expressed in linear units. More commonly one deals with 2-port devices of which only 3 are of relevance: the transmission line, attenuator, and amplifier. The ideal transmission line has an S-matrix, expressed in linear units:
\begin{equation}
    S=\begin{bmatrix}
        0 & 10^{\gamma l/10}\\
        10^{\gamma l/10} & 0
    \end{bmatrix}
\end{equation}
where $\gamma=\alpha+i\beta$, a complex propagation constant, and $l$ is the length of the line. One can see that the power attenuation in the line would be $\alpha l$ (in dB units), and the phase delay $\beta l$ of a given frequency would be related to the wavelength of the inputted frequency such that $\beta=10\ln(10)(2\pi/\lambda)$ in radians. A non-ideal transmission line would exhibit reflections, therefore non-zero reflection components; it could also be non-reciprocal, resulting in different propagation constants in each direction, $\gamma_1$ and $\gamma_2$. The ideal attenuator can be thought of as an ideal transmission line with a zero phase shift component:
\begin{equation}
    S=  \begin{bmatrix}
        0 & 10^{\alpha/10} \\
        10^{\alpha/10} & 0
        \end{bmatrix}
\end{equation}
where the $l$ has been absorbed into the attenuation value $\alpha$ and expressed in dB, and is entirely real and negative. non-ideal versions would exhibit some phase shift (imaginary component), or the same flaws as the transmission line. Similarly an ideal amplifier would be a further subset of the attenuator, only exhibiting amplification in the forward direction to the next port with gain, $G>0\, dB$:
\begin{equation}
    S=  \begin{bmatrix}
        0 & 0 \\
        10^{G/10} & 0
        \end{bmatrix}
\end{equation}
The only 3-port device covered here is the RF circulator which is used in the subsequent sections. Similar to the ideal amplifier, an ideal circulator would only allow for RF signals to travel in one direction exclusively to the adjacent port ($1\rightarrow2, 2\rightarrow3, 3 \rightarrow1$), in this case with zero loss or gain ($G=0\, dB$):
\begin{equation}
    S=\begin{bmatrix}
        0 & 0 & 1\\
        1 & 0 & 0 \\
        0 & 1 & 0
    \end{bmatrix}
\end{equation}
In reality, circulators aren't perfectly lossless, perfectly isolating, or perfectly frequency-independent, giving each circulator a unique set of S parameters along the 9 possible signal paths at each inputted frequency. Their deviation from the ideal can change with environmental conditions. For instance, to obtain their non-reciprocal nature in practice, circulators are made with ferrite; Incident microwaves interact with the static magnetic field of the ferrite and based on the direction they enter are attenuated or transmitted in different directions. This makes circulators inherently magnetic, and their performance sensitive to the high magnetic fields of an axion haloscope environment. In this way, the S-parameters can be thought of as a 'fingerprint' not only of the device, but the environment it finds itself within. In the next section, we will stress the unique challenges of the haloscope environment and the complexity of deciphering S-parameters in such a RF network.

\subsection{Cryogenic RF experiments}
\label{sec:CryoRF}
In the previous section, we outlined how several simply RF devices can be defined by their scattering parameters exactly in the ideal case. In reality, their practical construction and environment makes their S-parameters much more unique when measured experimentally on a network analyzer. Most commonly, all real RF devices are rated to perform to specifications only within a certain frequency range; This results in a unique signature on the VNA when a larger frequency range is swept out. 
\par The cryogenic environment can have a drastic effect on the performance of devices, often requiring specially manufactured components in order to maintain the desired RF specifications: For instance, a cryogenic attenuator tested for $10\,dB$ at $4\, K$ might have a drastically different attenuation at room temperature or even $40 \, K$. 
\par Although manufacturers try their best to create versatile, robust devices in these harsh conditions, even small deviations can be observable through the S-parameters. For instance, if a dilution refrigerator starts operating at a warmer temperature, heating all components uniformly the base noise level attenuation read in the RF receiver will be increased slightly, reflecting the increase in thermal Johnson noise. If only one component, such as an amplifier, is heated, this could uniquely change the S-parameters for the system, in this case by changing the gain profile of the amplifier. 
\par In reality, a change in fridge performance would adapt components uniquely according to the temperature gradients within the fridge. When one considers that there can be hundreds of connections and components in a RF network within a cryostat, it is clear that a human operator cannot diagnose the complex system of changes within its S-parameters throughout the duration of an experimental run. Nonetheless, if changes to S-parameters are unique and repeatable for given conditions, it might be possible for a different type of intelligence to learn to predict these phenomenon over time.
\subsection{Artificial Neural Networks}
\label{sec:ANNs}
Artificial Neural Networks (NNs) could very well be the solution to diagnosing a myriad of phenomenon within the cryogenic RF environment. NNs are learnable non-linear maps between arbitrary dimensional data input space to an arbitrary target space. NNs are made of composable parameterzied functions called 'layers'; modern automatic differentiation libraries have built-in methods to update parameters to minimize error functions between an initially random output space to the target space in a process called deep learning. Although it maybe unexpected to an observer, a mapping could exist between collected data and physical changes in the system that a neural network can be be optimized to recognize.  In the context of axion haloscopes, these trained networks could be used to alert operators to abnormal physical conditions or device performance.

\subsection{Principal Component Analysis}
In this work we will couple the expressive power of NNs with principal component analysis (PCA), a common tool used to reduce high dimensional spaces to lower dimension \cite{PCAReview}. For any centered data-sample matrix with n samples, $X$, the data-covariance matrix can be defined as $\frac{1}{n}X^{\top}X$. The eigenvectors (a linear combination of the initial data-fields) of this data-covariance matrix are known as the principal components of variance, and one can choose to represent the initial data matrix in the basis of these eigenvectors of variance. The corresponding eigenvalues provide information on the explained variance in the total dataset of each individual eigenvector, and so data scientists commonly truncate their basis by removing elements with low total explained variance.

While truncating the basis is necessarily throwing information away, there are in fact many benefits to applying PCA: first, higher-dimensional input spaces require higher dimensional parameterized layers, and in the regime of the NNs this work utilizes, this allows us to avoid overfitting and reduce training time. Second, because the eigenvectors are orthogonal to each other the resulting final feature space will not exhibit any collinearity between features. Finally, many signals in high dimensional data are actually low dimensional signals embedded in a noisy high dimensional space, and it is this noise we are aiming to remove when we perform PCA.

\subsection{Extracting diagnostic information using from scattering parameters}
\label{sec:extractdiagnostics}
This paper will begin by analyzing a simplified, non-cryogenic and non-superconducting experiment composed of a single RF circulator and RF cylindrical cavity. We use simple supervised machine learning methods and consumer PyBoard micro-electronics motors to change the state of our experiment, measure the resulting scattering parameters, and learn the mapping between these scattering parameters and the physical changes. While rudimentary, our first two experiments are a proof-of-concept and important stepping stone for the last experiment. In this final experiment, we we record VNA scans and thermometer reads of a cryogenic RF circulator undergoing dilution refrigerator warm-up cycle. We show how the temperature of the RF circulator can be regressed from features in the VNA scan using artificial intelligence.

\section{Machine Learning for Device Characterization}
In this section we detail our main experiments in order of increasing complexity. In subsection~\ref{sec:exp1} we demonstrate that a neural network in conjunction with PCA can learn to recognize circuit components by their scattering parameters. While simple, the learning framework developed for recognizing components is re-used for the remainder of the experiments. In subsection~\ref{sec:exp2} we show that NNs can learn the mapping between scattering parameters and physical quantities; specifically, we learn labels coupled to the external magnetic field strength around an RF circulator due to a permanent magnet. Finally, in subsection~\ref{sec:exp3} we learn the mapping between network scattering parameters of a cryogenic circulator device and resistance measured across ad ruthenium-oxide temperature probe mounted to the circulator in a dilution refrigerator as it undergoes a cool down. We aim to show how across these different levels of complexity, PCA and neural networks are able to extract meaningful features from scattering parameters to perform useful tasks.

\subsection{Experiment 1: Boolean device characterization}
\label{sec:exp1}

To begin, we start with an experiment to train a NN to recognize which of two RF circulators are the device under test to a VNA. While simple, we use this experiment to demonstrate our general method to extract features from scattering parameters and unambiguously show a task success. Because we are detecting which of two circulators are the device under test, this task is a binary classification problem and our performance can be understood through the classification accuracy metric. 

\subsubsection{Methods}
Our apparatus consists of a Keysight E5063A vector network analyzer (VNA), a minicircuits RC-8SPDT-A8 switchboard, a PC equipped with a NVIDIA\texttrademark Titan X graphical processing unit, with 12 GB of VRAM, and two identical LNF-CIC4\_12A circulators \footnote{https://lownoisefactory.com/product/4-12-ghz-single-junction-isolator-circulator/} whose s-parameters are indistinguishable by the naked eye, when viewed on a VNA. 

\begin{figure}[!ht]
    \centering
    \includegraphics[scale=0.55]{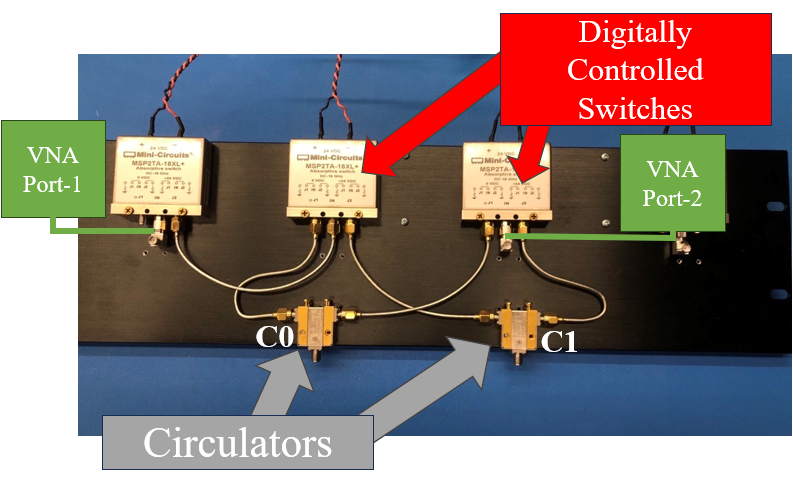} 
    \caption{\small\textbf{Binary Classification Apparatus} Two circulators (bottom), labeled C0 and C1, are wired to a digital switch board such that a VNA can measure the transmission coefficient (S21) through the first and third port of each circulator depending upon the switch state. The second port of each circulator is left open and unterminated. This was done in order to have the injection signal transmit through more of the circulators' pathways, so that the resultant signal would capture more of the unique characteristics of the given circulator. The wiring is such that the ports are kept the same between the VNA and circulator pairs to minimize any differences in the S21 parameter not originating from the circulators.}
    \label{fig:SetupExp1}
\end{figure}

We begin by wiring the circulators to the digital switchboard as shown in \ref{fig:SetupExp1}. This allows us to reliably switch in each circulator as the device under test to the VNA, also controlled digitally via the PC. We collect data by randomly switching in either circulator, recording the switch state and scattering parameters with the VNA. The VNA was set to collect 1,000 frequency points linearly sampling the region 5.5 Ghz to 12.5 Ghz. We collected with an input power of -15 dBm, with an IBW of 300 kHz. All collections were taken sequentially on the same day.

After collections completed, we separated the dataset out into a random 0.75/0.25 training and test split (2000 VNA scans in the dataset). The test split data is held out from the training process and is used to evaluate the performance of the algorithm. This is a standard procedure in the AI/ML field to prevent effects like memorizing/over-fitting on the training data influencing our perception of the model's performance.

We limit ourselves to the features of the S21 trace given that the S21 path would be a transmission mode measurement through each port of the circulators. Observing that there was a bias between the mean value of the S21 trace for our circulators, we whiten the data to remove this trend. This bias is likely due to the differences in the wiring paths rather than differences in the circulators themselves. Regardless of the source, removing the bias makes the problem more interesting as we now are relying on our feature extraction pipeline to learn something about the 'fingerprint' of the circulators on the S21 trace. Because of the high correlation between the traces, we used principal component analysis to represent the 2000 point traces using the first ten principal values. These first 10 PCs make up only 6.7\% of the total variance. 

Next, we train a NN to learn the mapping from the PCA feature vector to the switch state labels. Our neural network architecture is a feed forward network with 1 hidden layer of size 512 and hyperbolic tangent activation functions \cite{activationfunctions}. We train the neural network on the binary classification task by minimizing the binary cross entropy loss using the stochastic gradient descent algorithm. We perform all AI experiments in the PyTorch framework \cite{PyTorch}.

\subsubsection{Results}
The neural network is able to classify which circulator is the device under test with high accuracy (100\% on the testing data set). To probe why the neural network is able to classify to such high accuracy, we plot the scatter plot of PCA\_dim0 vs PCA\_dim1 in figure~\ref{fig:exp1} below. Observe that a vertical line could be drawn at PCA\_dim0 = 0.0 that would perfectly separate the clusters formed from the embeddings of each component. This shows that the learning task is indeed very trivial since PCA has isolated a single feature that separates our classes. In this context, we should not be surprised that the neural network performs very well, as the problem is simple. Nonetheless, this demonstrates how in practice wide-band measurements of scattering parameters can be used as a fingerprint to characterize RF networks.

Owing to the fact we can easily describe a good solution along PCA\_dim0 = 0, we can evaluate whether our AI aligns with this solution. If the network learned a decision boundary around PCA\_dim0 = 0, then the remaining 9 feature values would essentially be irrelevant. While we do not evaluate this hypothesis analytically, we will show that random samples drawn from a simplified feature space follow this solution. 

We create a synthetic dataset by sampling from a normal distribution parameterized by the mean and variance of each of the training principal components. We then replace the feature of PCA\_dim0 and PCA\_dim1 from a grid of points on the interval [-5,5] and [-5,5], sampled every 0.1 in value. This synthetic dataset assumes that there are no covariance between features; however, since these features are the result of PCA this assumption is satisfied. 

Using the neural network output on these synthetic datapoints we can create a heatmap of the neural network prediction confidence along the grid of PCA\_dim0 and PCA\_dim1 features. We underlay this heatmap in the PCA feature space of our points in figure~\ref{fig:exp1}, showing that the neural network indeed has a decision boundary along the vertical PCA\_dim0 = 0 line. This is a visual confirmation that the NN has learned the expected solution.

Finally, to evaluate whether the network is learning the wiring differences or learning the differences between the circulators themselves, we replace the physical circulators in our switchboard with each other (but keep the labels the same). The right plot of figure~\ref{fig:exp1} demonstrates the results, where we observe that although there is a distribution shift from the original training data, the neural network retains a nearly perfect accuracy (99.7\%). This implies the PCA+NN system has learned to separate the circulator features from the remainder of the RF Network (e.g., wiring differences).   

\begin{figure*}[!ht]
    \centering
    \includegraphics[scale=0.35]{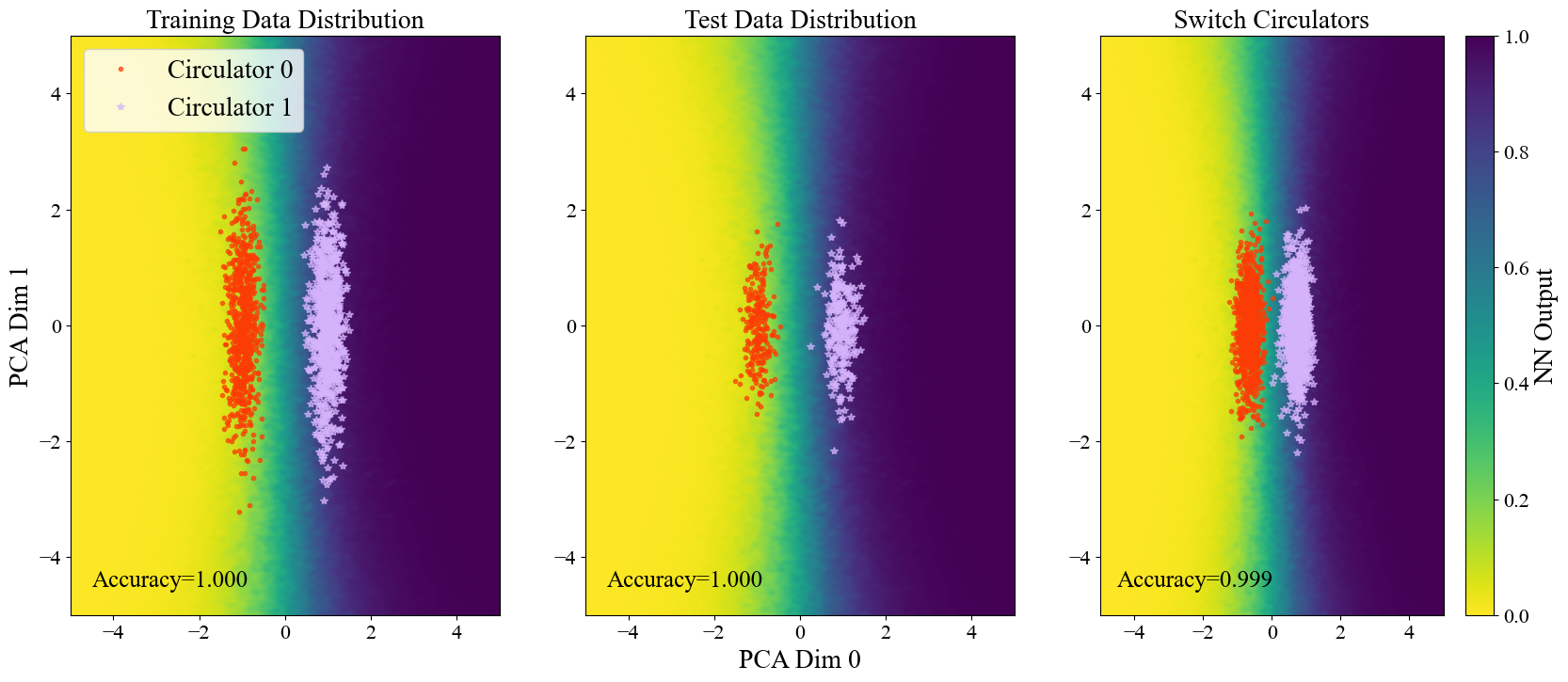} 
    \caption{\small\textbf{Binary Circulator Recognition via Scattering Params} We plot the distribution of PCA-reduced features from the VNA for the training data (left) the test data (center) and a second evaluation set formed by switching the wiring paths of the physical circulators (right). The background heatmap shows the approximate prediction confidence of the NN in the PCA feature space. We note that the confidence transitions in prediction around vertical line along PCA dim0 value of 0, which approximately is the mid-point formed between class clusters in the training data; which is entirely expected.}
    \label{fig:exp1}
\end{figure*}

\subsection{Experiment 2: Regressing Environmental Magnetic Field Strength}
\label{sec:exp2}
Our goal in this section is to demonstrate that the scattering parameters can be used to monitor the external environment in RF-networks while two degrees-of-freedom vary. As motivation, recall that a real axion haloscope experiment operates a large magnet around an RF cavity to induce a non-zero $\vec{B} \cdot \vec{E}$ product. The remainder of the read-out chain is composed of sensitive superconducting amplifiers that must be shielded from this magnetic field, achieved in ADMX through a second, opposite oriented, solenoid \cite{ADMXSQUIDs}. If this sensitive system fails, there are only two hall probes to detect the change in magnetic field, and these sensors are limited by their position and orientation \cite{ADMXDesign}. Therefore, it can be difficult for an operator to recognize a fault. In this section we ask whether off-band monitoring of scattering parameters in a mock reflection mode measurement could be used to recognize changes in magnetic field around a circulator, e.g., to the ends of monitoring excess magnetic fields and alerting operators to a fault.

\begin{figure*}[!ht]
    \centering
    \begin{minipage}{0.25 \textwidth}
    \includegraphics[width=\textwidth]{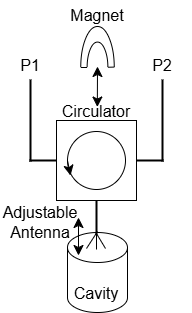} 
  \end{minipage}
    \begin{minipage}{0.65 \textwidth}
    \includegraphics[width=\textwidth]{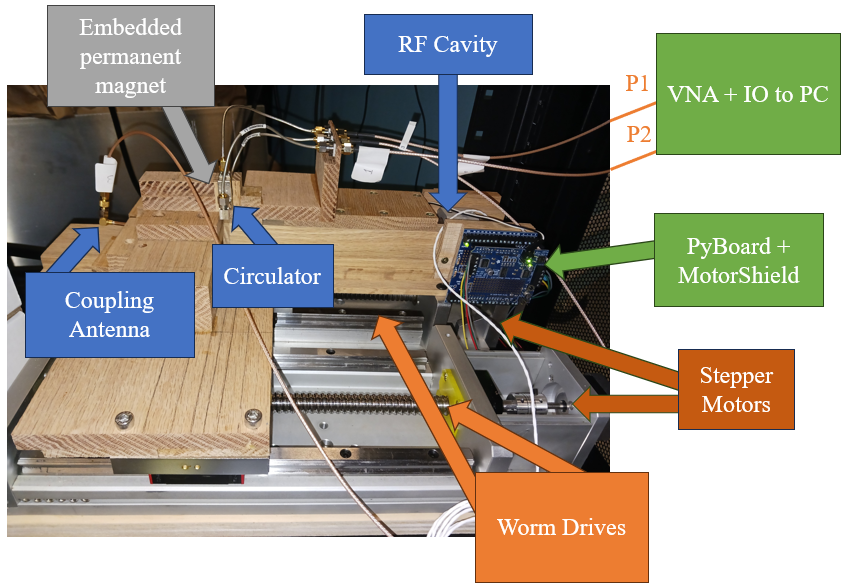} 
  \end{minipage}
    \caption{\small\textbf{Two Degree-of-Freedom Experiment Apparatus} Two stepper motors controlled by a PyBoard micro-controller are connected to separate worm-drives. The first motor controls the distance between a permanent magnet and a circulator, while the second motor controls the insertion depth of a coupling antennae to a RF cavity. These motors are controlled separately, and therefore two degrees-of-freedom are achieved. Using the VNA, we make a cavity reflection measurement through the RF circulator: Port 1 (P1) injects a signal into the first port of the circulator that reflects off the coupling antennae inside the RF cavity (this is connected to the adjacent circulator port), passing back through the circulator, and then is measured through port 2 (P2) that is connected to the final port of the circulator.}
    \label{fig:SetupExp2}
\end{figure*}

\subsubsection{Methods}
To study the use of out-of-band RF measurments to probe magnetic field impact on device response, we modify our experiment apparatus. using a Pyboard Metro M4 express and Arduino Motor Shield, we control two identical Adafruit NEMA-17 stepper motors\footnote{https://www.adafruit.com/product/324} as the two degrees of freedom. The first motor drives a coupling antenna into a cylindrical RF resonant cavity, which mimics the coupling antennae control that axion haloscope experiments must manage to read-out the axion signal \cite{ADMXDesign}. The second motor adjusts the position of a neodymium permanent magnet relative to the circulator, which changes the external magnetic field around the device. We show a picture of the experimental apparatus in Figure~\ref{fig:SetupExp2}.

The magnetic field induced by a permanent magnet can be modeled as a magnetic dipole. Recall in the far-field approximation for a magnetic dipole, the field strength goes as $B \propto \frac{1}{R^3}$.\cite{GriffithsEM} In this experiment, we will satisfy our goal by in principle allowing the magnetic field strength to be completely determined by the label $R$. This is done for 1) ease as measuring $R$ is much easier than measuring the external magnetic field strength directly, and 2) performance as typically we would like evenly and linearly sampled targets for our regression task\cite{ExperimentalDesign}.

Our VNA port 1 (P1) is wired to port 1 of the Circulator, port 2 is wired to the resonant cavity coupling antennae, and port 3 is wired to port 2 (P2) of the VNA. Therefore, in this case, an S12 measurement measures a reflection off the coupling antennae inside the resonant cavity. We perform a data collection by stepping sequentially through a grid of positions that linearly sample the magnet motor space from 0 to 10,000 steps in 10 step increments and the coupling antennae insertion depth space from 0 to 10,000 steps in 100 step increments. The magnet motor '0' step position corresponded to the magnet's closest position to the circulator without contact, whereas the antenna motor '0' position corresponded to the antenna maximally inserted into the cavity; 1000 steps of both motors is approximately 2.5 cm travel of the worm drives. The position data was then scaled to a range of 0 to 1 in both dimensions for input into the neural network.

We repeat the same PCA feature extraction and NN training pipeline as described in subsection~\ref{sec:exp1}, though in this experiment we use all scattering parameter amplitudes and keep the first 20 principal components of each trace for a total of 80 input features.

\subsubsection{Results}
\begin{figure*}[!ht]
    \centering
    \includegraphics[scale=0.8]{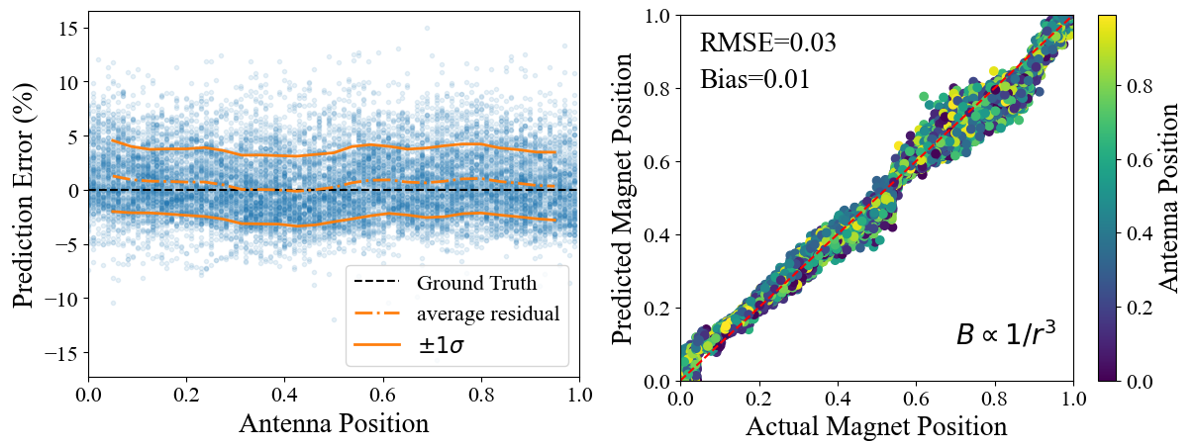} 
    \caption{\small \textbf{Left:} The percentage residual error of between the model's prediction and the true magnet's position plotted against the antenna motor position. While it is biased to over predict the true value slightly, the residuals have little covariance with the antennae position, meaning that the model is homoskedastic in performance across Antennae position. \textbf{Right} A second visualization of the NN performance as a point-for-point plot, where the predicted magnet position is plotted against the true magnet position. The diagonal red dashed line is parity, and the color of the scatter points are scaled with the Antennae position.}
    \label{fig:Tukey2}
\end{figure*}

We plot the predicted magnet position against the true position in Figure~\ref{fig:Tukey2}, which shows a tight correlation between the prediction and real values ($R=0.98$), a bias in the residuals of $\text{Bias}=0.03$, and a average scatter in the residuals as $\text{RMSE}=0.04$. The prediction performance is largely invariant to the insertion depth, which can be more readily visualized by a Tukey plot \ref{fig:Tukey2} \textbf{Left}. The correlation between residual and antennae position is ($R=0.016$, with p-value of $p=0.06$, meaning we do not reject the null hypothesis that R=0) , which is evidence that our NN predictions of the magnet position are (linearly) invariant to the Antennae position. The takeaway from this demonstrations is that the environment, (in this case, a magnetic field), may impart changes to device s-parameters and those changes can be learned (and later identified) by AI. 



\subsection{Experiment 3: Regressing device temperature in a changing cryogenic environment}
\label{sec:exp3}

In our third experiment, we evaluate the same PCA+NN methodology for a device embedded in a dilution refrigerator cooled to cryogenic temperatures. The goal was to use the scattering parameters of the device to predict the temperature state of the fridge. This represented another leap in complexity for the NN prediction; the cool-down process of a dilution refrigerator is not precisely controllable. This is in contrast to the movement of the magnet that was controllable and could produce an evenly sampled data set in section \ref{sec:exp2}. In this case, the distribution of samples would be uneven based on the rate of cooling at a given time.

\subsubsection{Methods}
We mounted a LNF-CIC4\_12A circulator to the mixing chamber (MXC) stage of a BlueFors LD400 dilution refrigerator. The circulator is the device under test for our VNA. Under the nomenclature of section \ref{sec:scattering_params}, port 1 of the VNA was connected to the first port of the circulator, port 2 of the VNA was connected to the second port, and the third port was left open. VNA scans were taken periodically every 5 minutes during the fridge cooling process. Concurrent with the VNA, the temperature of each of the 4 temperature stages ($50\,\mathrm{K}$, $4\,\mathrm{K}$, still, and MXC plates) were measured by ruthenium oxide temperature sensors and logged by a Lake-shore\texttrademark resistance bridge roughly every minute over the cool-down process, which lasted roughly 5 days. Because the temperature was logged more often than the VNA measurements, the temperature data was interpolated at the times of the VNA measurements, which allowed a temperature value of each stage to be associated with a VNA measurement. 
\par NNs would then be trained to predict the temperature of a given stage based on the circulator S-parameters; note that a network was trained for each of the four stages and each produced a single temperature prediction rather than a single network that produced four temperature predictions simultaneously.
\subsubsection{Results}
\par
\begin{figure*}[!ht]
    \centering
    \includegraphics[scale=0.8]{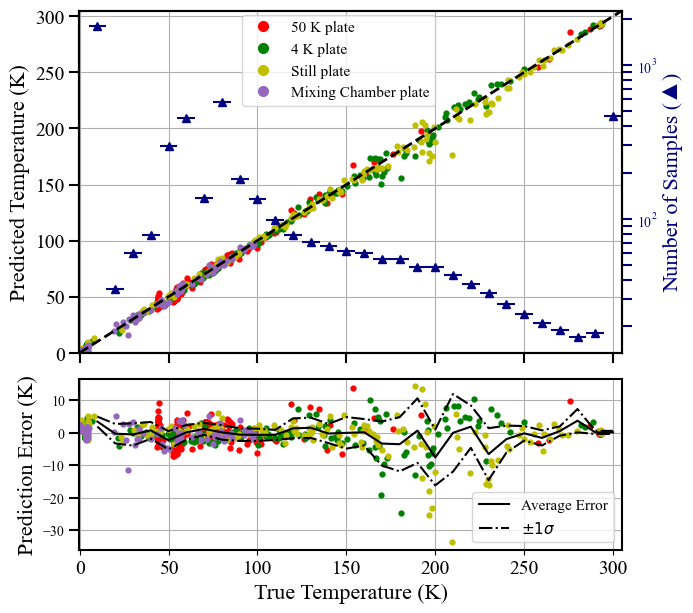} 
    \caption{\textbf{Top:} The performance of the NN learning from calibrated temperature labels visualized with a plot of NN predicted versus the true temperature recorded by the corresponding temperature stage sensor. Each point represents a single test prediction, and is colored by the temperature stage of the fridge the NN was trained to predict. The mixing chamber stage sensor is only calibrated to read below temperatures of $100 \,K$, while the other three stages have values from $0$ to $300\, K$. Additionally, the total number of test samples per $10\, K$ versus temperature is shown as triangle markers corresponding to the second y-axis in log scale. \textbf{Bottom:} The absolute prediction error expressed in Kelvin for every test point in each of the four temperature stage test sets versus true temperature. The total average prediction error across all four temperature stage NNs versus the true temperature is shown as well as the $\pm 1 \sigma$ lines. These were calculated by binning all samples for every $10\, K$ from $0$ to $300\,K$ and calculating the mean and standard deviation for each bin set.}
    \label{fig:exp3}
\end{figure*}

\begin{figure}[!ht]
    \centering
    \includegraphics[scale=0.7]{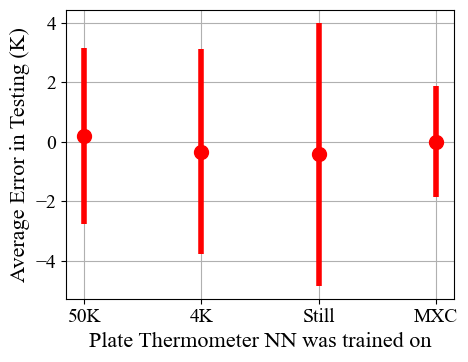} 
    \caption{The average prediction error for each of NNs trained for a given fridge plate temperature stage across the entire temperature range of the fridge cool-down.}
    \label{fig:exp3_avgerr}
\end{figure}

 The final results of this training exercise are shown in Fig. \ref{fig:exp3}. The prediction error at a given temperature was, unsurprisingly, inversely correlated with the number of test samples taken at that temperature. Presumably if one cooled at a slower, fixed rate in the $150$ to $250 \, \mathrm{K}$ range, taking as many measurements per Kelvin  as the $50$ to $100 \, \mathrm{K}$ range, the higher prediction error seen in that region would decrease to the lower values seen in the $50$ to $100 \, \mathrm{K}$ region. Each NN for all four temperature stages were able to predict the temperature within $\pm\, 5\,\mathrm{K}$ on average (Fig. \ref{fig:exp3_avgerr}, with the NN trained to predict the mixing chamber stage (MXC) temperature performing the best with an average error of $0\pm 1.9 K$. This is most likely because the MXC stage had a high sample density across its testing range from $0$ to $100\, K$ while the other stages included samples taken in lower sample density regions. 

\par During the analysis process of creating these NNs several parameters were optimized to mitigate prediction error.
\par The network was originally trained to use all 4 scattering parameters as inputs to the NN. However, it was noticed, unsurprisingly, that certain S-parameters contained significantly less information and features related to the circulator, and would actually increase the prediction error in the NN; For instance, $S_{12}$ data would be useless because it involved sending a signal backwards through the circulator, which would be highly attenuated and lack any consistent features, hence it was removed as an input. An analysis was conducted on which combination of S-parameter inputs resulted in the lowest prediction error, and this was found to be the $[S_{22},S_{11}]$ combination. In our final results (shown in Fig. \ref{fig:exp3} and \ref{fig:exp3_avgerr}), only the $S_{22}$ and $S_{11}$ parameters were used as inputs for the four NNs trained for each temperature stage.
\par
\begin{figure}[!ht]
    \centering
    \includegraphics[scale=.6]{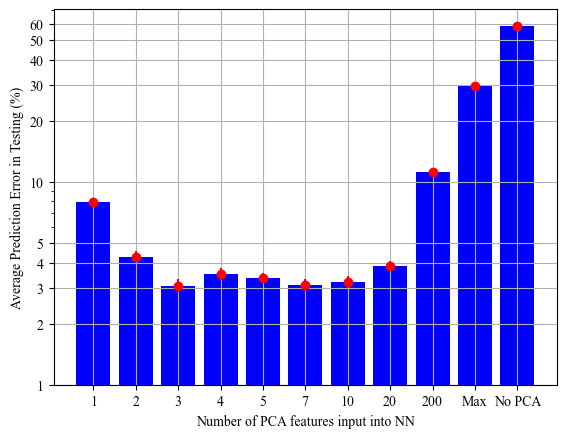} 
    \caption{The average percentage prediction error using varying amounts of PCA features as inputs to the NN. In this case, the NN was trained for prediction of the $ 50\, K$ plate temperature. The $S_{11}$ and $S_{22}$ complex parameters were reduced by the same number of PCA features listed across the x-axis. The 'Max' label refers to a PCA feature input number equal to the number of samples within the training set ($N=1128$). The 'No PCA' label refers to a NN with no PCA pre-processing, that takes every VNA point (1600 points per sweep) as an input into the NN.}
    \label{fig:exp3_PCAerr}
\end{figure}

 Additionally, the use and optimization of PCA was found to be crucial for getting accurate predictions. PCA was used to reduce the data of each scattering parameter trace into a subset of features, then combine each VNA trace's principal components into a single reduced feature vector. It was found that the explained variance of the first 5 features was roughly about 86-90\% of the total variance in the $S_{11}$ data and 27\% in the $S_{22}$ data. This meant that including additional features as inputs to the given NN did not give it that much more information for more precise identification, and actually had the effect of increasing prediction error as shown in Fig. \ref{fig:exp3_PCAerr} for the $50 \, K$ plate and reducing each S-parameter by the same amount of features. When the PCA pre-processing was not used at all, and instead took each individual VNA point value as an input to the NN, the average error was nearly 60\% in the case of Fig. \ref{fig:exp3_PCAerr}. Similarly, the maximum amount of PCA features was used ($N=1128$ samples in the training dataset), the error was roughly $30\%$ and often much higher for the other lower temperature stages. The minimum error across temperature stages and S-parameter input combinations always lay between a PCA feature number of 2 to 20, and therefore it was settled to use 5 features in the final NN results shown in Fig. \ref{fig:exp3}.

\section{Discussion}
\label{sec:Discussion}
Our goal was to demonstrate that VNA scans can essentially behave as a kind of fingerprint for physical phenomena occurring in an RF network and that those fingerprints may perhaps be learned by AI to aid in In situ characterization of RF experiments.  We have shown in three different experiments that scattering parameter scans can be reduced using PCA and neural networks to predict a change in circuit components and predict changes in the environment including temperature and external magnetic field strength. We interpret this initial success as a green light to investigate more complex usage of neural networks to interpret RF Networks. 
\par This work does not come without its limitations. In these 3 experiments, the networks were very simple and purpose driven to reduce the degrees of freedom in the training process. As shown in section \ref{sec:exp1}, these neural networks can pick up on patterns unique to specific devices, which may necessitate follow-up training any time a component is changed within the RF network. Although we show that a neural network is capable of learning specific physical changes despite another degree of freedom being varied (Section \ref{sec:exp2}), this may not always be the case as more degrees of freedom are added or if that DoF is itself effected by the 'labeled' degree of freedom in the training set.  In section \ref{sec:exp3}, the network relies on training with a well calibrated temperature sensor that is thermalized with the RF device under testing, and any deviation from that accuracy will be carried over into the model. However, all these limitations can be improved upon with future study. 
\par This work points the way to several follow-up experiments for future work. In the case of experiment 1, one would want to expand this identification ability to be able to recognize specific components across different common RF network layouts; the output line of the cold ADMX receiver vs. the arrangement in a BlueFors fridge of a collaborating institution for instance. Similar to experiment 2, work is in progress training a NN to predict the location of a tuning rod within a copper cavity based on its wide-band S parameters, independent of the antennae position; this application has proved more challenging because of the sensitivity of the cavity S-parameters to rod position distances that are on the order of the 'backlash' in the stepper motors themselves. This necessitates a better calibrated tuning system both in training and testing. Future work implementing the 'Noisy Sci-kit RF' open software package developed at PNNL has been proposed; perhaps a neural network could be trained on simulated S-parameters that could then subsequently predict behavior about the real physical system. Future work is also being pursued to automate the calibration process of quantum-limited amplifiers in order to maximize their gain.
\section{Conclusion}
\label{sec:Conclusions}
This work demonstrated that neural networks could predict and characterize three simple RF networks based on their wide-band scattering parameters as inputs. The first experiment demonstrated that S-parameters can act as a 'fingerprint' for specific devices. The second experiment showed that predictions can be made about a variable in the devices' physical environment, in this case its proximity to a magnetic field, despite unrelated variables also changing within the training set, the antenna insertion depth in this experiment; this was done with a well curated training dataset. Finally, this idea was taken to the cryogenic environment, where the contents of the training dataset couldn't be as well controlled: a thermal cycling of a dilution refrigerator.  Nonetheless, a neural network could predict the temperature of an RF component during the cycle through the wide-band scattering parameters. We hope this work inspires more complex applications of machine learning in the axion detection field. As axion haloscopes increase in complexity, these types of AI controls will help alleviate the mystery when diagnosing problems in such systems for the human operators. 
\begin{acknowledgments}
 PNNL is a multi-program national laboratory operated for the U.S. DOE by Battelle Memorial Institute under Contract No. DE-AC05-76RL01830. This work was supported by the DOE Office of Science, High Energy Physics, C. Boutan Early Career Award (FWP 77794 at PNNL). This research made use of scikit-rf, an open-source Python package for RF and Microwave applications.
\end{acknowledgments}

\bibliography{bibliography}

\appendix


%
%

%

\end{document}